# Irreducible sub-nm$^2$ ferroelectric domains by 2D-flat polar band in perovskite superlattices

**Pawan Kumar*[1] and Jun Hee Lee*[,1,2]**

[1]Department of Energy Engineering, School of Energy and Chemical Engineering, Ulsan National Institute of Science and Technology (UNIST), Ulsan, 44919, Republic of Korea

[2]Graduate School of Semiconductor Materials and Devices Engineering, Ulsan National Institute of Science and Technology (UNIST), Ulsan, 44919, Republic of Korea

*Corresponding author. Email: pawankumar@unist.ac.kr and junhee@unist.ac.kr

**Abstract**

Flat polar phonon bands in ferroelectrics have been envisioned to realize irreducibly small domains for achieving the highest memory density. However, such bands are extremely rare, with only a one-dimensional flat polar band discovered in ferroelectric hafnia, which gives rise to irreducibly narrow, line-type domains with a half-unit-cell width of 2.7 Å. Here, we report the discovery of a two-dimensional flat polar band in the ferroelectric $(BaTiO_3)_1/(BaXO_3)_1$ superlattice (X = Zr, Sn), which generates extremely localized, noninteracting dipoles within its quarter-unit-cell area that form quasi-degenerate states of irreducible sub-nm$^2$ ferroelectric domains. The estimated ferroelectric domain density, exceeding 300 $Tbit/cm^2$ (assuming one bit per domain) represents a record high among the known ferroelectrics, with independently switchable dipoles at low voltages. The ferroelectric phase hosting two-dimensional flat bands constitutes a competing ground state across superlattice stacking directions and thus can be experimentally realized in both freestanding structures and epitaxial growth on MgO substrates. This discovery opens unprecedented opportunities to explore multidimensional flat polar bands in ferroelectrics and to engineer ultra-dense and low-power memory devices.

## Introduction

Ferroelectric memories[1] offer technological advantages such as fast read/write speeds, low power consumption, and high endurance, making them highly promising for artificial intelligence[2], neuromorphic computing[3], and next-generation nanoelectronics[4]. They utilize the intrinsic properties of ferroelectric domains, whose spontaneous polarization can be switched by an external electric field. Information is encoded in ferroelectric domains[5], regions with distinct polarization directions which are separated by domain walls (DWs). Each domain in a ferroelectric corresponds to one bit, and thus smaller domains enable higher memory density. However, achieving smaller ferroelectric domains remains challenging owing to dispersion in a polar phonon band. In conventional ferroelectrics, the polar phonon mode is softer than any other modes in the dispersive band and the soft polar phonon[6] requires finite-size domains[7] (at least a few $nm \sim 100\ nm$) to condense to produce polarization, resulting in quite low memory densities (0.01~0.1 $Tbit/cm^2$)[8–10]. The flatness of the polar bands is therefore essential, as it inherently stabilizes ultranarrow domains whose widths are a fraction of the unit cell, making the domain wall energy (DWE) nearly equal to that of the bulk state.

The flat polar bands, however, are extremely rare in ferroelectrics, with only a single case of a one-dimensional flat polar band discovered in $HfO_2$[11], which produces the narrowest one-dimensional domains with nearly zero DWE, partially addressing the dense-memory challenge owing to the limitation of a one-dimension flat band. Therefore, the lack of two-dimensional flat bands in ferroelectrics continues to hinder the realization of ultra-dense square-shaped memories. In contrast, two-dimensional flat electronic bands, analogous to polar phonon bands, have already been observed in several materials and exhibit remarkable exotic properties[12,13]. For example, two-dimensional flat electronic bands appear in frustrated spin systems[14] and twisted bilayer graphene[15], producing highly degenerate ground states, and display unconventional magnetism and superconductivity. While the discovery of two-dimensional flat polar bands remains elusive, extensive theoretical and experimental efforts continue to pursue their realization.

Although the unstable polar bands in the cubic phase of $BaTiO_3$[6], a material widely used in ferroelectric devices[16], are less dispersive[6] than those in other perovskite ferroelectrics, they still destabilize the narrow DWs in its ferroelectric tetragonal phase. In recent years, superlattice engineering has become a widely used approach for tailoring material properties due to their interfacial couplings[17–22]. Ferroelectric superlattices, fabricated via layer-by-layer deposition of functional thin films, have been extensively studied and revealed unique phenomena, including hybrid improper ferroelectricity in perovskites and fluorite oxides[17,22], stabilization of charged DWs[23,24], and the realization of negative capacitance[25]. Moreover, leveraging the structural similarities between ferroelectric $BaTiO_3$, paraelectric $BaZrO_3$ and $BaSnO_3$, the $(BaTiO_3)_n/(BaXO_3)_m$ ($BT_n/BX_m$) (X = Zr, Sn) superlattices (SLs) have already been extensively studied[26–31].

In this manuscript, using symmetry analysis and interface engineering of superlattices, we strategically mix polar and antipolar modes, which results in a robust two-dimensional flat polar band in the ferroelectric state of the columnar-ordered $BT_1/BX_1$ SL. Our first-principles simulations demonstrate that these two-dimensional flat polar bands produce highly localized dipoles within an exceptionally small area of sub-nm², precisely 0.31 (0.33) $nm^2$, resulting in an ultra-high ferroelectric domain density of ~320 (~303) $Tbit/cm^2$ in the ferroelectric $P4mm$ phase of $BT_1/BZ_1$ ($BT_1/BS_1$) SL, which translate into ultra-high-dense memory. These dipoles are intrinsically localized and independently switchable within the two-dimensional plane, giving rise to the narrowest domains with a vanishingly small DWE of 2.2 (1.5) $mJ/m^2$ in the $BT_1/BZ_1$ ($BT_1/BS_1$) SL, yielding a manifold of quasi-degenerate states of ferroelectric domains. Remarkably, the polar $P4mm$ phase of both superlattices stabilizes as the ground-state structure in the columnar ordering. Their dynamic stability further supports the feasibility of experimental realization, either through single-crystal growth or epitaxial thin-film growth.

## Results

### Underlying mechanism of a 2D-flat polar phonon band

We begin our analysis by examining the unstable polar $\Gamma_{15}^{z}$ ($\Gamma_4^-$) mode (**Fig. 1 and Table S1**) in the high symmetry paraelectric cubic $Pm\bar{3}m$ phase of pristine $BaTiO_3$, which reveals the dispersive nature (**Fig. 1a and S1a**) of the polar band in momentum space. The condensation of this unstable polar mode in the reference $Pm\bar{3}m$ phase transforms it into the ferroelectric $P4mm$ phase (**Fig. S1b and c**). The positive curvature of the polar band (**Fig. 1a**) contributes to the DWE[32] cost and, because the paraelectric–ferroelectric phase transition in $BaTiO_3$ involves only a single mode, the polar band remains dispersive in the ferroelectric $P4mm$ phase (**Fig. S1d**), preventing the stabilization of narrow DWs.

To flatten this polar band in two-dimensional momentum space, we employed strategic polar-antipolar mode mixing, and found that the equal mixing of polar $\Gamma_{15}^{z}$ and antipolar $M_3'^{z}$ ($M_3^-$) modes (**Fig. 1b**) of the cubic phase of pristine $BaTiO_3$, which are also connected to the same band (**Fig. 1a**), cancels the antiferroelectric displacements while enhancing the ferroelectric ones (**Fig. 1b**). This mixing gives rise to ferroelectric and spacer regions in the 2D plane, called ferroelectric columns (FC) and spacer columns (SC), respectively, in bulk $BaTiO_3$ (**Fig. 1b)**. In this arrangement, only a quarter of the unit cell becomes polar, while the rest remains non-polar, which can potentially store one bit per unit cell. Group-theoretical and symmetry analysis revealed that the mixing of the $\Gamma_{15}^{z}$ and $M_3'^{z}$ modes can be mediated by an antipolar $M_4$ ($M_4^+$) mode (**Fig. 1b**) of the cubic phase of pristine $BaTiO_3$, which exhibits a trilinear coupling ($\Gamma_{15}^{z} M_4 M_3'^{z}$) and could potentially condense in a new ferroelectric state in $BaTiO_3$. However, the strength of this coupling is insufficient to simultaneously condense $M_3'^{z}$ and $\Gamma_{15}^{z}$ modes in pristine $BaTiO_3$. Consequently, this mixed polar–antipolar ferroelectric state is unlikely to stabilize, as its energy increases continuously with the amplitude of these modes under constrained condensation (**Fig. S2**). When the constraint is removed, the mixed state relaxes back to the ferroelectric $P4mm$ phase, where only the polar $\Gamma_{15}^{z}$ mode condenses. Moreover, the $M_4$ mode is intrinsically hard (**Fig. 1a**) and unlikely to soften under external perturbations such as strain, which might otherwise strengthen the trilinear couplings that drive polar–antipolar mode mixing in the ferroelectric phase.

**Superlattice engineering to realize 2D flat polar bands**

Since the emergence of the $M_4$ mode is essential for mixing the $\Gamma_{15}^{z}$ and $M_3'^{z}$ modes, we designed two columnar-ordered superlattices[30], $BT_1/BZ_1$ and $BT_1/BS_1$, by stacking alternating layers of $BaTiO_3$ and $BaZrO_3$ ($BaSnO_3$) along the [1 1 0] direction (**Fig. 2a**), where Ti and Zr/Sn are alternately surrounded by each other within the xy-plane. Since Zr/Sn has a larger atomic radius than Ti, oxygen atoms in the same plane shift diagonally away in the xy-plane from Zr/Sn and toward Ti, generating an oxygen displacement pattern identical to that of the $M_4$ mode (**Fig. S3**). Thus, despite being the highest-frequency phonon mode in the $Pm\bar{3}m$ phase in $BaTiO_3$ (**Fig. 1a**), $BaZrO_3$ (**Fig. S4a**) and $BaSnO_3$ (**Fig. S4b**), a robust, insuppressible non-polar $M_4$ mode inherently emerges in the optimized $BT_1/BZ_1$ and $BT_1/BS_1$ SLs which gets the $P4/mmm$ space group (**Fig. 2a**). This interfacial engineering provides a natural pathway for activating the $M_4$ mode and enabling the required mixing of polar-antipolar modes, which is otherwise inaccessible in bulk $BaTiO_3$. Notably, similar interfacial engineering has been employed to enhance phonon–phonon couplings in pristine $PbTiO_3$ and $HfO_2$, where the resulting $PbTiO_3/SrTiO_3$[17] (**see SI-note 1**) and $HfO_2/CeO_2$[22] SL exhibit improper and hybrid-improper ferroelectricity, respectively. Thus, the discovery of a two-dimensional flat polar band induced by superlattice engineering in ferroelectrics introduces a new ingredient into superlattice-engineered ferroelectricity.

The phonon spectra of $P4/mmm$ phase of the columnar ordered $BT_1/BZ_1$ (**Fig. S7a.**) and $BT_1/BS_1$ (**Fig. S7b.**) superlattices exhibit only a single polar instability, in contrast to the triply degenerate unstable polar modes in pristine $BaTiO_3$ (**Fig. 1a**), where two instabilities vanish due to columnar-ordering of ferroelectric-paraelectric superlattices. The unstable polar $\Gamma_2'^z$ ($\Gamma_2^-$) mode involves atomic displacements along the z-direction (**Fig. 2 and S7c**), and its decomposition into the modes of cubic $BaTiO_3$ confirms that the $\Gamma_{15}^z$ and $M_3'^z$ modes are equally mixed ($\Gamma_2'^z = \Gamma_{15}^z + M_3'^z$) (**Fig. S7c-e**). We note that the polar $\Gamma_2'^z$ mode originates from the $BT_1/BZ_1$ superlattice, whereas the $\Gamma_{15}^z$ and $M_3'^z$ modes originate from pristine $BaTiO_3$. This mixing results in a remarkably flat polar unstable band in the Brillouin zone within the 2D-plane along the $\Gamma - X - M - \Gamma$ path (**Fig. S7a and b**) of both superlattices. The frequencies of the $BT_1/BZ_1$ SL at the high-symmetry points show excellent agreement with its previously estimated phonon frequencies[30].

The condensation of the polar $\Gamma_2'^z$ mode in the $P4/mmm$ phase of the $BT_1/BZ_1$ ($BT_1/BS_1$) superlattice transforms it into a new polar $P4mm$ phase (**Fig. 2b and S8**). We note that ferroelectric tetragonal phase in pristine $BaTiO_3$ also have the same space group. The estimated polarization (**using eq. 1**) of 25.3 (21.1)$\mu C/cm^2$ confirms the ferroelectric nature of the $BT_1/BZ_1$ ($BT_1/BS_1$) superlattice. Surprisingly, despite containing 50% of the non-polar $BaZrO_3$ ($BaSnO_3$) component, the polarization decreases by only ~27 (~38) % compared to the ferroelectric tetragonal $P4mm$ phase of pristine $BaTiO_3$ (34.5 $\mu C/cm^2$). This indicates that polarization is not a simple average of the paraelectric and ferroelectric constituents but rather reflects the robust presence of ferroelectricity. Furthermore, the polarization in the $BT_1/BZ_1$ ($BT_1/BS_1$) superlattice is concentrated within the Ti centric quarter of the unit cell (**Fig. S9**), referred to as the ferroelectric column, while the remaining 75% of the unit cell, the spacer column, remains non-polar. Consequently, the Ti-centric ferroelectric column exhibits a local polarization of 101.2 (84.4) $\mu C/cm^2$.

The phonon spectra of the ferroelectric $P4mm$ phase of $BT_1/BZ_1$ (**Fig. 2c**) and $BT_1/BS_1$ (**Fig. 2d**) superlattices reveal no imaginary frequencies, confirming their respective dynamical stabilities. Interestingly, their respective polar phonon bands exhibit remarkable flatness throughout the Brillouin zone in the xy-plane, indicating the presence of highly localized dipoles in this plane. This exceptional flat polar band enables an extensive manifold of degenerate polar states, allowing individual dipoles to switch within the xy-plane. Thus, the $BT_1/BZ_1$ and $BT_1/BS_1$ superlattices lead to the localization of dipoles in columns surrounded by spacer regions in the xy-plane, which surprisingly results in the narrowest square domains within a quarter-unit-cell area. In contrast, in the ferroelectric $Pca2_1$ phase of $HfO_2$ (**Fig. 2e**), the polar band is flat only along one direction (**Fig. 2f**), which restricts dipole localization to lateral layers and results in line-type ferroelectric domains. Moreover, the $P4mm$ phase is found to be the ground state in both $BT_1/BZ_1$ and $BT_1/BS_1$ superlattices (**Table S2 and S3**), indicating that domain formation is not influenced by competing lower-energy structures.

**Stability and mobility of ultrahigh-density ferroelectric domains**

To demonstrate the stability of the narrowest square-shape domains in the $BT_1/BZ_1$ ($BT_1/BS_1$) superlattice, we constructed DW configurations by flipping a single dipole within a 4×4×1 supercell of the $P4mm$ phase (**Fig. 3a**). Remarkably, the single-dipole-switched structures relaxed into stable configurations with a vanishingly small DWE of 2.2 (1.5) $mJ/m^2$, indicating quasi-degenerate states with the bulk and revealing the weakly-interacting nature of individual dipoles, an inherent feature of flat-band ferroelectricity. The switched unit cell retains a substantial polarization of −19.8 (−18.5) $\mu C/cm^2$ (**Fig. 3b and c**), comparable to the bulk value, confirming that no significant polarization loss occurs during DW formation. The estimated single dipole switching barrier of 0.135 (0.083) $eV/uc$, obtained from nudged elastic band (NEB) simulations (**Fig. 3d**), is comparable to the Landau switching barrier of 0.137 (0.085) $eV/uc$, as both processes share the same intermediate $P4/mmm$ phase. Furthermore, we performed single-dipole switching simulations in the ferroelectric $P4mm$ phase of pristine $BaTiO_3$ (**Fig. S10a and b**). However, these DWs destabilize, and the system spontaneously relaxes into the rhombohedral $R3m$ ground state, where the polarization aligns along the body-diagonal [111] direction and the z-component of polarization in the switched unit cell reverses (**Fig. S10c and d**). This transformation arises because the ferroelectric $P4mm$ phase in $BaTiO_3$ is metastable, and the local lattice distortions induced by domain formation generate strain fields that drive the system into the $R3m$ ground state.

To investigate domain-wall mobility in the $BT_1/BZ_1$ ($BT_1/BS_1$) superlattice, we considered two successive dipole reversals: the first from the bulk state (**Fig. S11a**) to a single-dipole switching configuration (**Fig. S11b**), followed by the reversal of its adjacent dipole (**Fig. S11c**). The estimated domain-wall energy (DWE) of 2.2 (0.8) $mJ/m^2$ in the adjacent-dipole-reversal configuration is nearly identical to that of the single-dipole-switched structure, while the flipped polarization in the adjacent unit cell is −19.7 (−18.8) $\mu C/cm^2$, confirming that the two-adjacent-dipole-reversal state is well stabilized with full polarization retention. The estimated energy barrier of 0.131 (0.081) $eV/uc$ (**Fig. S11d**) indicates that each dipole experiences nearly the same barrier during reversal, consistent with the weakly-interacting behavior of dipoles. This weakly-interacting nature in both $BT_1/BZ_1$ and $BT_1/BS_1$ superlattices implies that reversing any individual dipole yields the quasi-degenerate states, with a polarization magnitude comparable to the bulk.

From an experimental perspective, the system can be generalized to an N×N×1 supercell (with N unit cells along x and y), allowing up to $N^2$ quasi-degenerate polarization configurations, each separated by identical switching barriers (**Fig. 4a**), which are also comparable to their respective Landau switching barriers (**Fig. 4b**). Despite having comparable energy barriers, local switching becomes overwhelmingly dominant, as its rate scales with the large number of available configurational sites ( $\Gamma_{local} \propto N^2 \, exp(-\Delta E/k_B T)$ ), in contrast to the Landau switching ($\Gamma_{Landau} \propto exp(-\Delta E/k_B T)$), making local switching both kinetically and statistically favored. Thus, each unit cell of $BT_1/BZ_1$ ($BT_1/BS_1$) superlattice can store a stable ferroelectric bit that can be reliably read and written, reaching the ultimate physical limit of ferroelectric memory under ambient conditions. With a unit cell area of ~0.31 (~0.33) $nm^2$, this translates to an ultimately high domain density of ~320 (~303) $Tbit/cm^2$. Interestingly, modern nano-sized electrodes[33,34] can precisely control local switching to target selected dipoles (**Fig. S12**), allowing memory architectures to be scaled down to unprecedented dimensions, dramatically reducing the fabrication footprint and enabling ultra-dense ferroelectric memory. Ideally, a well-controlled periodic electric-field pulse applied to thin-film samples can generate polarization plateaus whose number approaches $N^2$, enabling deterministic multilevel operation.

Moreover, the estimated Landau/local switching barrier of 1.0 (0.63) $meV/Å^3$ for the $BT_1/BZ_1$ ($BT_1/BS_1$) superlattice is higher than (or similar) that in pristine $BaTiO_3$ (0.60 $meV/Å^3$) (**Fig. S13**). The estimated energy barrier for homogeneous polarization switching in $BaTiO_3$ is in very good agreement with previously reported theoretical values[35], supporting the accuracy of our DFT simulations. Since the coercive field ($E_c$) scales linearly with the ratio of the switching barrier ($\Delta E$) to polarization ($E_c \propto \Delta E/P$)[36], $E_c$ in $BT_1/BZ_1$ ($BT_1/BS_1$) is predicted to be only 2.3 (1.7) times higher than in $BaTiO_3$. Considering the experimental $E_c$ of ~2 $kV/cm$ in bulk $BaTiO_3$, both superlattices are expected to exhibit $E_c$ in range of 4 to 5 $kV/cm$, significantly low for practically feasible polarization switching, although the values can vary with film thickness and sample defects[37].

**Route to the experimental realization of a $BT_1/BX_1$ superlattice**

We have now examined the stability of the $BT_1/BX_1$ (X = Zr, Sn) columnar-ordered superlattices relative to the rocksalt-ordered along the [111] direction (**Fig. S14a**) and the layered superlattice along the [001] direction (**Fig. S14d**). The rocksalt-ordered phase does not exhibit any paraelectric-to-ferroelectric transition in the $BT_1/BX_1$ SL, as its paraelectric $I4mmm$ phase itself is the lowest energy state, in agreement with earlier first-principles results[30], which is further confirmed by the absence of imaginary modes in its phonon dispersion (**Fig. S14b and c**). In the layered-ordered superlattices, the phonon spectra of the paraelectric $P4/mmm$ phase of $BT_1/BX_1$ (**Fig. S14e and f**) exhibits the highest instability in the polar mode (**Fig. S14g**). The condensation of this unstable polar mode transforms it into a polar $Pmm2$ phase with polarization along the x direction (**Fig. S15**); however, this phase remains energetically higher than the columnar superlattice in both cases (**Table S2 and S3**). Interestingly, the $I4mmm$ phase in rocksalt ordering is only 0.83 (1.66) $meV/atom$ lower in energy than that of ferroelectric columnar-ordered $BT_1/BZ_1$ ($BT_1/BS_1$) superlattices (**Table S2 and S3**), indicating that the rocksalt-ordered paraelectric $I4mmm$ and columnar-ordered ferroelectric $P4mm$ phases in the $BT_1/BX_1$ superlattice stabilize at similar energy scale. This suggests that, in bulk $BT_1/BX_1$, the ferroelectric columnar ordering can exist as a competing ground state at room temperature. Thus, our atomic-scale simulations confirm that the bulk ferroelectric state in free-standing columnar-ordered $BT_1/BX_1$ SL can be experimentally realized using single-crystal growth techniques and can be integrated between two electrodes for memory devices (**Fig. 4c**).

We further extended our analysis to the epitaxial growth of columnar $BT_1/BZ_1$ and $BT_1/BS_1$ thin films. We found that the lattice parameters (**Table S2 and S3**) of the ferroelectric phase of both superlattices closely match those of the cubic phase MgO (a= 4.184 Å), a common substrate for growing BT/BZ superlattice thin-film structures[38,39]. Since vertically aligned superlattice growth is experimentally challenging, we considered a layer-by-layer horizontal stacking (**Fig. 4d**), with polarization oriented in-plane. Using the in-plane lattice parameters of the $\sqrt{2} \times \sqrt{2} \times 1$ supercell of MgO substrate, we optimized the out-of-plane lattice parameter and atomic positions of the columnar $BT_1/BZ_1$ ($BT_1/BS_1$) superlattice. Remarkably, the estimated polarization of 25.3 (23.5) $\mu C/cm^2$ remains bulk-like, as in-plane polarization is largely unaffected by epitaxial strain, unlike out-of-plane polarization, which is highly strain-sensitive[40]. These results further confirm the robust ferroelectricity of the $BT_1/BX_1$ superlattices and their insensitivity to epitaxial growth conditions, highlighting their technological potential for device applications, as the film can be placed between two lateral electrodes (**Fig. 4d**).

**Discussion**

In summary, we have discovered a 2D-flat polar band in the $BT_1/BZ_1$ and $BT_1/BS_1$ superlattices by strategically mixing polar and antipolar modes, enabling the ultrahigh memory density. This 2D flat polar band allows the formation of highly localized, weakly-interacting dipoles confined to a plane, theoretically achieving an ultrahigh density of ferroelectric domains exceeding 300 $Tbit/cm^2$, hundreds of times higher than that of current ferroelectric materials. The stabilization of the polar state within sub-nm² regions, dominated by a single atomic displacement and with polarization switching that can be selectively controlled by an external electric field, also enables the development of atomic-scale semiconductors for ultra-high-density memory. Surprisingly, the ferroelectric state in the columnar-ordered $BT_1/BZ_1$ and $BT_1/BS_1$ superlattices exhibits competing ground states among all possible orientations, making its experimental realization feasible. This breakthrough could fundamentally transform the field, leading to ultra-high-density ferroelectric memory devices and pushing memory technologies far beyond current limits.

## Methods

### Computational details

The first-principles calculations in this study were performed using density functional theory (DFT) within a plane-wave pseudopotential framework as implemented in the Vienna *Ab initio* Simulation Package (VASP)[41–43]. The projector augmented-wave (PAW)[44] method was employed with the generalized gradient approximation (GGA) and the PBEsol[45] functional for the exchange-correlation energy. The valence electronic configurations considered were $5s^2 5p^6 6s^2$ for Ba, $3p^6 3d^2 4s^2$ for Ti, $4s^2 4p^6 4d^2 5s^2$ for Zr, $4d^{10} 5s^2 5p^2$ for Sn, and $2s^2 2p^4$ for O. The plane-wave energy cutoff was set to 500 $eV$. For Brillouin zone integrations, Monkhorst–Pack (MP)[46] k-point meshes of 8×8×8 and 6×6×8 were used for the $BaTiO_3$ primitive cell (5 atoms) and the $BT_1/BZ_1$ and $BT_1/BS_1$ superlattices (10 atoms), respectively. Structural optimizations were performed until the total energy converged to within $10^{-7}$ $eV$ and the residual forces were below $10^{-3}$ $eV/Å$. Lattice dynamical properties of $BaTiO_3$ and the $BT_1/BZ_1$ and $BT_1/BS_1$ superlattices were obtained using density functional perturbation theory (DFPT) as implemented in VASP.

### Estimation of polarization

The polarization in different ferroelectric phases was obtained by multiplying the Born effective charges ($Z^*$) of each atom with its displacements ($d$) from the centrosymmetric reference phase as follows,

$$P_s = \frac{e}{V} \sum_i Z_i^* . d_i, \qquad (1)$$

where $e$ and $V$ represent the electronic charge and supercell volume, respectively, while $i$ denotes the atomic site in the supercell.

**Data and materials availability:** All data supporting the findings of this study are available within the paper and its Supplementary Information.

**Author contributions:** JHL supervised this work. PK carried out the DFT calculations. PK and JHL developed the theory and wrote the manuscript. All the authors have read and approved the manuscript.

**Competing interests:** The authors declare no competing financial or non-financial interests.

## Acknowledgements

This study was funded by Basic Research Laboratory (Grant No. RS-2023–00218799), Nano & Material Technology Development Program (Grants No. RS-2024–00404361, No. RS-2024–00444182), Grants No. RS-2025-24535610 and No. RS-2023–00257666 through the National Research Foundation of Korea (NRF) funded by the Korea government (MSIT). This work was also supported by Industrial Technology Innovation Program (Grant No. RS-2025–06642983) and the Korea Institute for Advancement of Technology (KIAT) grant funded by the Korea Government (MOTIE) (Grant No. P0023703, HRD Program for Industrial Innovation), and the National Supercomputing Center with supercomputing resources including technical support (Grants No. KSC-2022-CRE 0075, No. KSC-2022-CRE-0454, No. KSC-2022-CRE0456, No. KSC-2023-CRE-0547, and No. KSC-2024-CRE-0545).

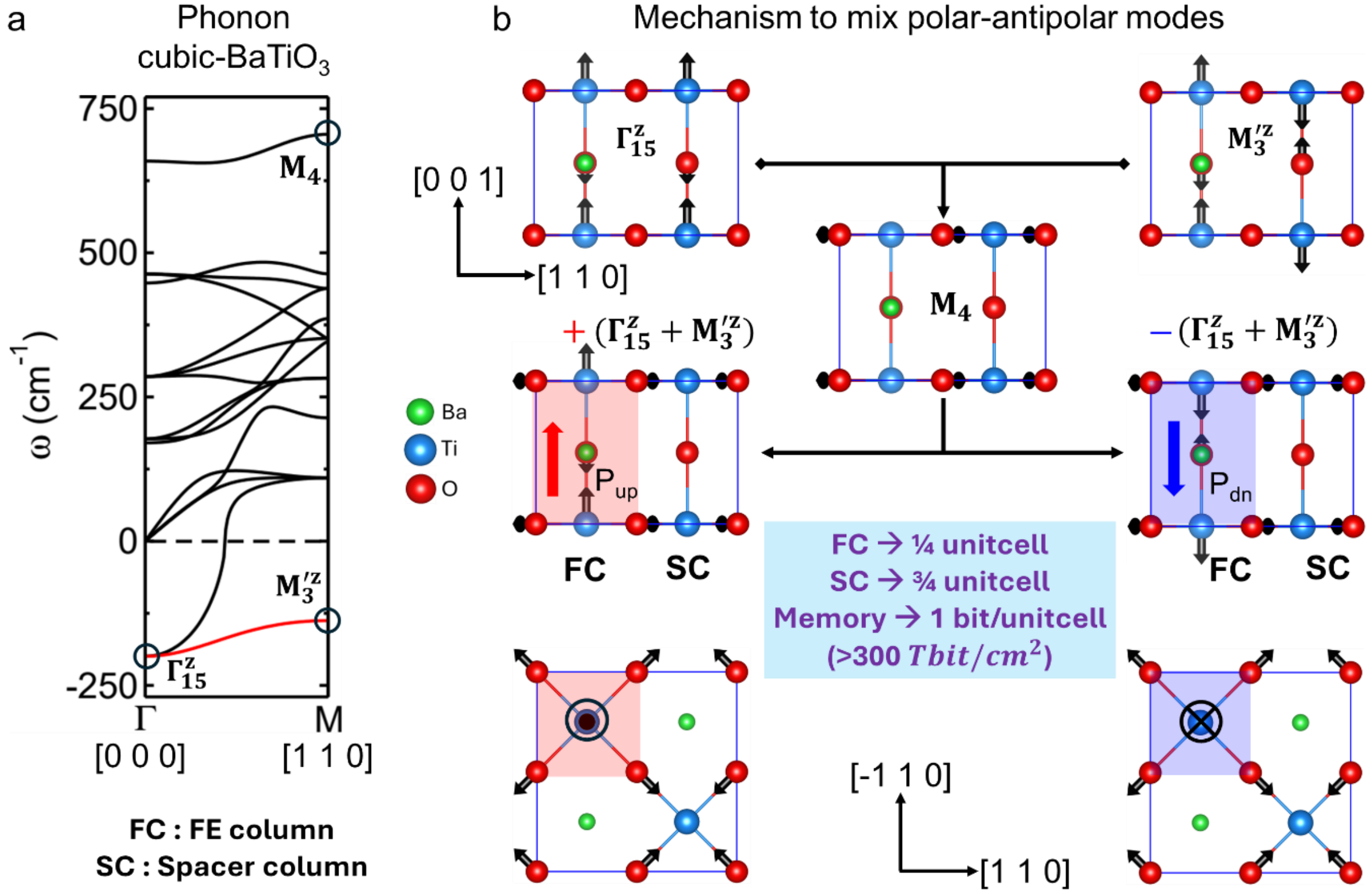

**Fig. 1. Mechanism for mixing of polar and antipolar modes in pristine $BaTiO_3$.** (a) Phonon dispersion of cubic $BaTiO_3$ along the Γ→M, showing a dispersive polar band, whose polar mode is the most unstable and is connected to the antipolar $M_3^{\prime z}$ mode. (b) Atomic displacement associated with the polar $\Gamma_{15}^{z}$ (top left), antipolar $M_3^{\prime z}$ (top right), and the highest frequency $M_4$ (middle) modes that exhibit trilinear coupling. Equal mixing of $\Gamma_{15}^{z}$ and $M_3^{\prime z}$ modes suppress the antipolar displacements, enhances the ferroelectric distortion, and generates symmetrically identical up (red shaded) and down (blue shaded) polar phases. In these ferroelectric phases, the polarization is confined within quarter of the unit cell, while the remaining unit cell remains nonpolar.

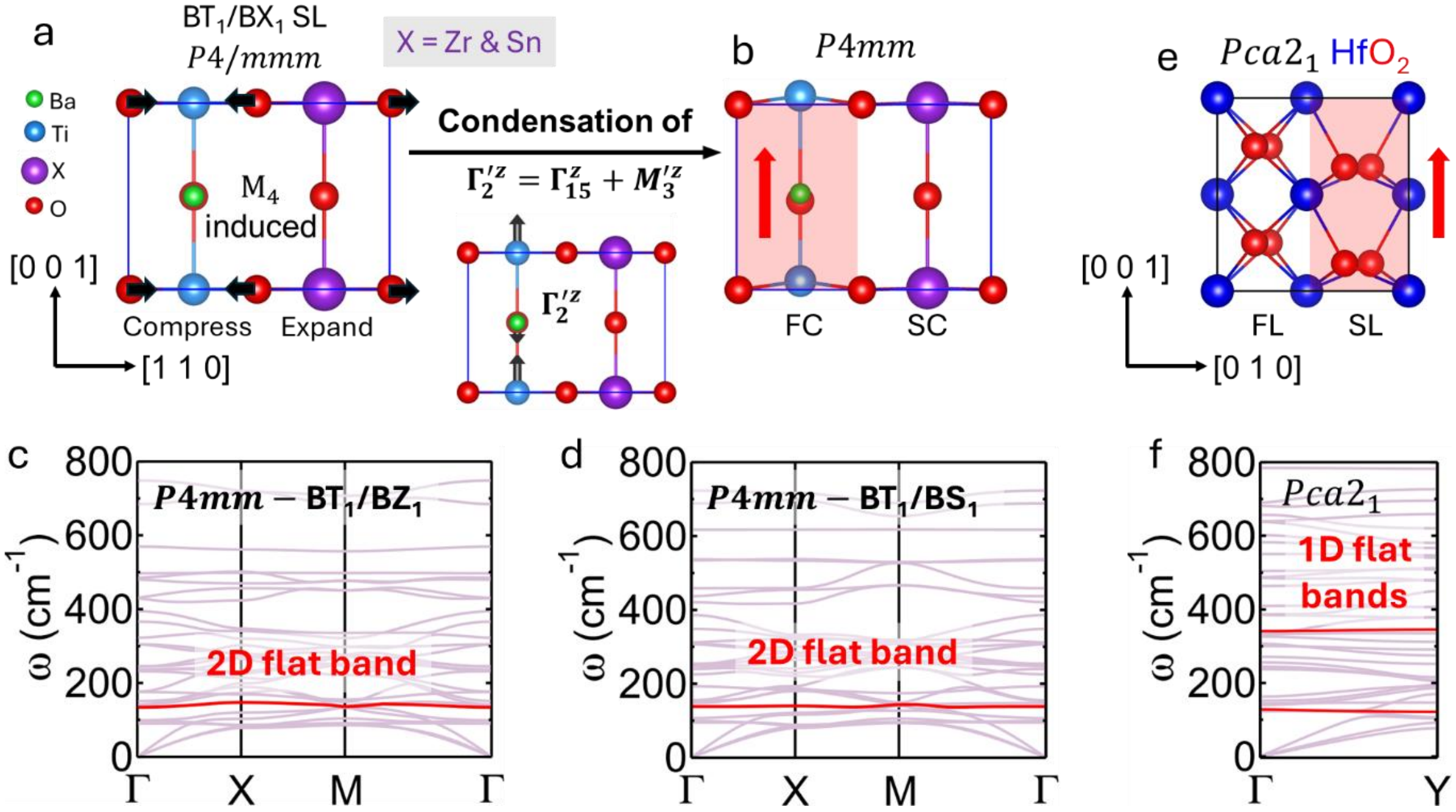

**Fig. 2. First-principles results of 2D-flat polar bands in $BT_1/BZ_1$ and $BT_1/BS_1$ SLs and their comparison with previously known 1D flat bands in $HfO_2$.** (a) Relaxed atomic structure of the centrosymmetric $P4/mmm$ phase of the $BT_1/BX_1$ superlattice, showing oxygen atoms displaced toward Ti and away from X due to the atomic size difference between Ti and X. The condensation of unstable polar $\Gamma_2'^z$ mode transforms $P4/mmm$ into the polar $P4mm$ phase (b), FC and SC denote ferroelectric and spacer columns, respectively. The phonon spectra of $P4mm$ phase of $BT_1/BZ_1$ (c) and $BT_1/BS_1$ (d) exhibit the remarkable two-dimensional flat band in momentum space with no imaginary frequencies, confirming their dynamical stability. (e) Atomic structure of the ferroelectric $Pca2_1$ phase (SL and FL denotes spacer and ferroelectric layers, respectively), whose phonon dispersion (f) exhibits only one-dimensional flat polar bands.

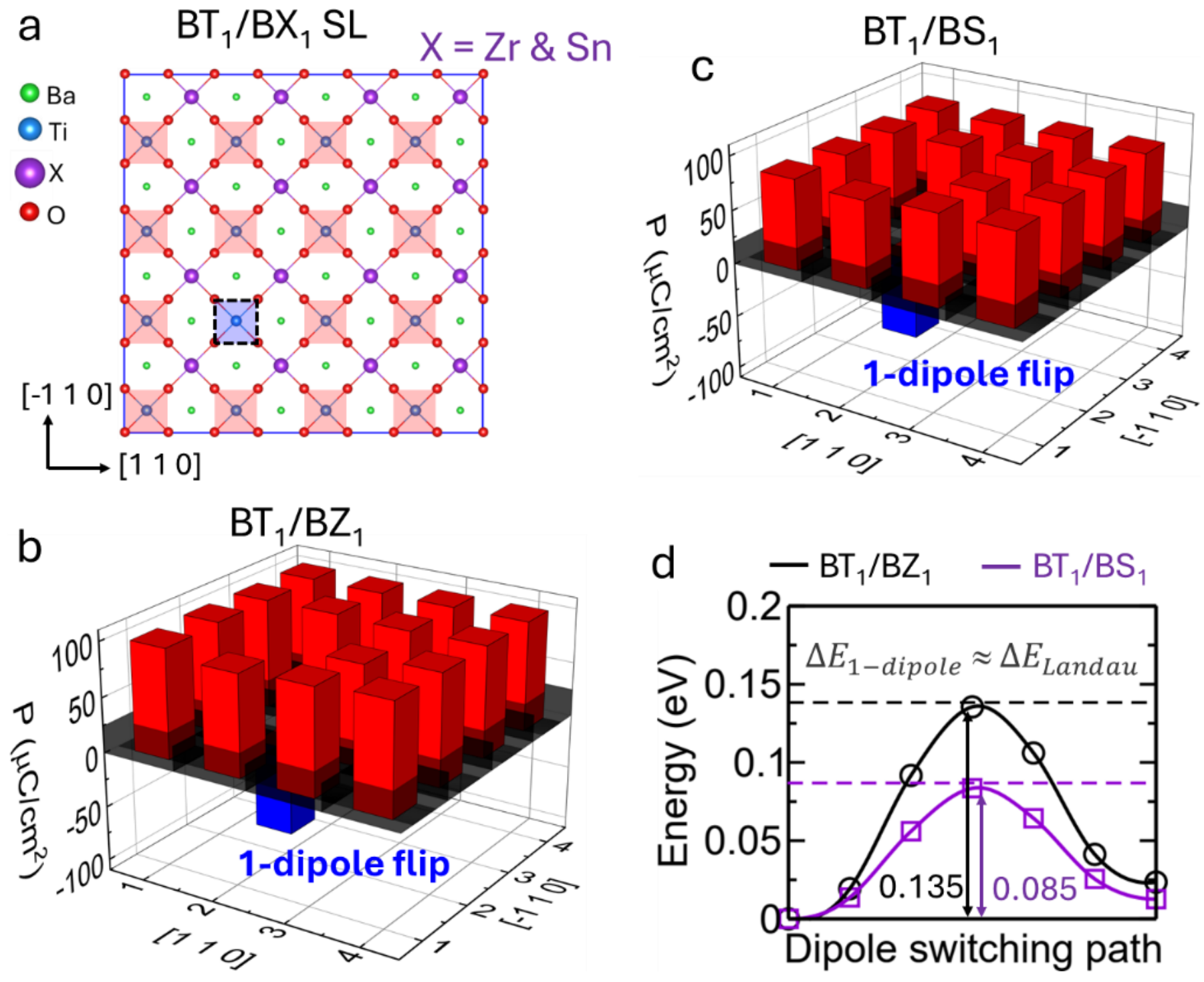

**Fig. 3. Results of individual polarization switching in the $BT_1/BX_1$ superlattice and their comparison with $HfO_2$.** (a) Atomic structure of a 4×4×1 supercell of $P4mm$ phase with one flipped dipole, where Ti-centered red squares denote up polarization, blue square indicate down polarization and white space depicts the nonpolar region. Average polarization (black) and localized polarization within one-quarter of the unit cell for the up-polarized (red) and down-polarized (blue) colomns in $BT_1/BZ_1$ (b) and $BT_1/BS_1$ (c) superlattices. (d) Energy landscapes associated with single-dipole switching in $BT_1/BZ_1$ (black) and $BT_1/BS_1$ (purple), with their respective homogeneous switching barriers indicated by black and purple dashed lines.

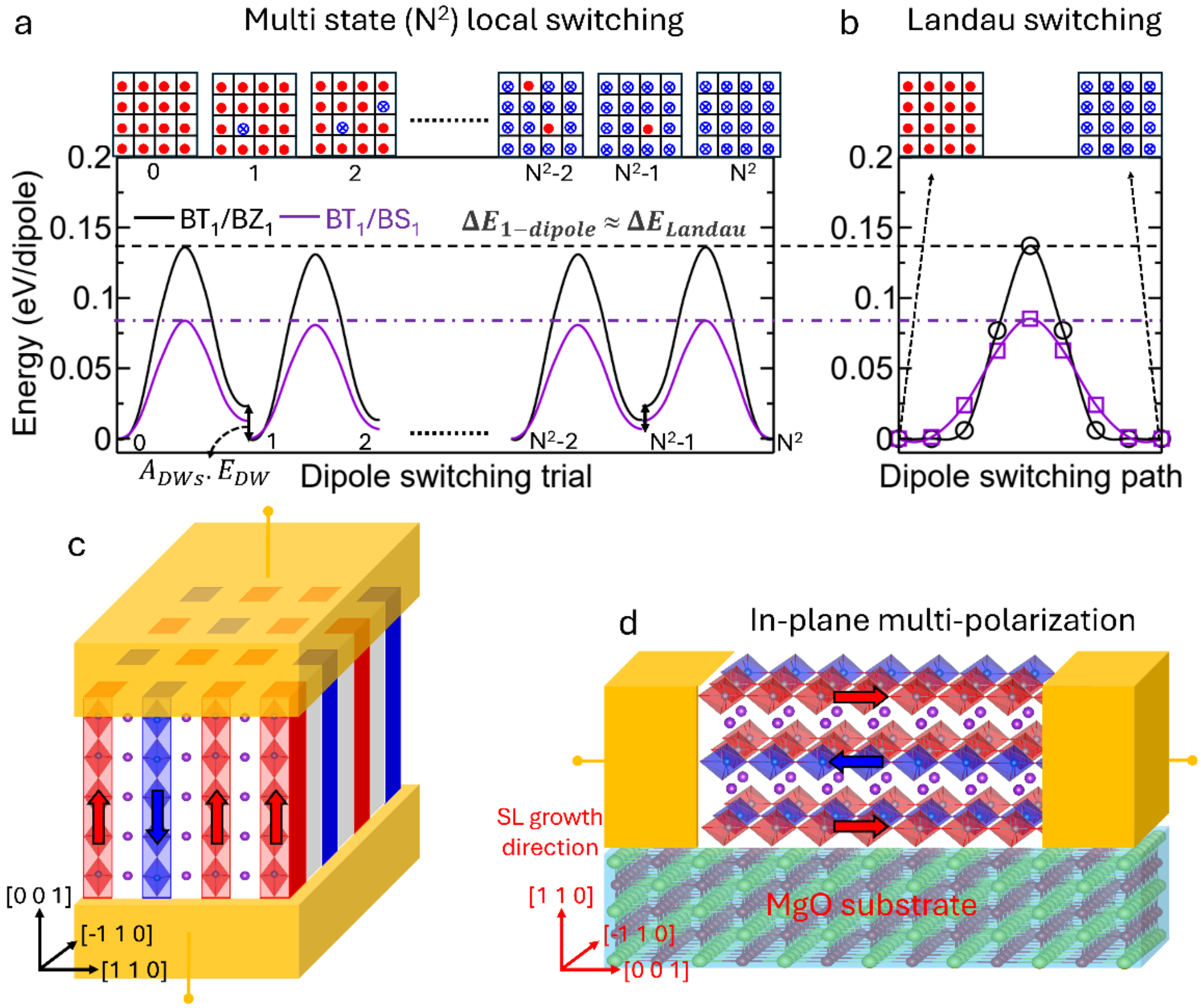

**Fig. 4. Schematic illustration of dipole switching in the $BT_1/BZ_1$ and $BT_1/BS_1$ superlattices for potential memory device applications.** (a) Generalization to an N×N×1 supercell, where up to $N^2$ polarization states can form that are nearly degenerate and separated by identical energy barriers. The slight energy jump in the energy landscape results in a vanishingly small domain wall energy ($E_{DW}$) times domain walls area ($A_{DWs}$). (b) Energy landscape associated with Landau switching in $BT_1/BX_1$ superlattices. Black dashed and purple dash-dotted lines across panels a and b mark the Landau switching barriers. (c) Schematic of the $BT_1/BX_1$ superlattice sandwiched between two electrodes (yellow) in a free-standing bulk form with randomly oriented polarization (up- and down-polarizations in each unit cell are indicated by red and blue bars, respectively). (d) Schematic of the $BT_1/BX_1$ superlattice epitaxially grown on an MgO substrate by layer-by-layer deposition, with randomly distributed in-plane polarization in each unit cell and kept between two lateral electrodes. Polarization along [0 0 1] and [0 0 -1] in-plane directions is indicated by red and blue Ti-centric octahedra, respectively.